\documentstyle[eqsecnum,preprint,aps,epsfig,floats]{revtex}
\begin{document}

\tighten


{\flushright{MZ-TH/99-11}
\flushright{IRB-TH/2000-1}
\flushright{TSU HEPI/2000-01}
\flushright{}
}

\vglue 1.3cm

\begin{center} \begin{Large} \begin{bf}
Analyticity, crossing and the absorptive parts of the one-loop
contributions to the quark-quark-gluon gauge boson four-point function
\end{bf} \end{Large} \end{center}
\vglue 0.35cm

{\begin{center}
\vglue 1.5cm
J.G.\ K\"{o}rner$^*$
\parbox{6.4in}{\leftskip=1.0pc
{\it Institut f\"{u}r Physik, Johannes
Gutenberg-Universit\"{a}t, D-55099 Mainz, Germany} }\\
\vglue 1.0cm
B.\ Meli\'{c}$^{\dagger}$
\parbox{6.4in}{\leftskip=1.0pc
{\it Theoretical Physics Division, Rudjer Bo\v{s}kovi\'{c} Institute,
        HR-10001 Zagreb, Croatia} }
\vglue 1.0cm
Z.\ Merebashvili$^{\ddagger}$
\parbox{6.4in}{\leftskip=1.0pc
{\it High Energy Physics Institute,
Tbilisi State University, 380086 Tbilisi, Georgia} }
\end{center}}


\begin{center}
\vglue 1.0cm
\begin{bf} ABSTRACT \end{bf}
\end{center}

\parbox{6.4in}{\leftskip=1.0pc
Starting from the known one-loop result for the
$e^{+}e^{-}$-annihilation process $e^{+}e^{-}\stackrel{\gamma,Z}
{\longrightarrow} q\bar{q}g$ with massless quarks we employ analyticity
and crossing to determine the absorptive parts of the corresponding
one-loop contributions in Deep Inelastic Scattering (DIS) and in the
Drell-Yan process (DY). Whereas the
${\cal O}(\alpha_s^2)$ absorptive parts generate a non-measurable
phase factor in the $e^{+}e^{-}$-annihilation channel one obtains
measurable phase effects from the one-loop contributions in the
deep inelastic and in the Drell-Yan case. We compare our results with the
results of previous calculations where the absorptive parts in DIS and in
the DY process were calculated directly in the respective channels.
We also present some new results on the dispersive and absorptive
contributions of the triangle anomaly graph to the DIS process.
}

\renewcommand{\thefootnote}{*}
\footnotetext{e-mail: koerner@thep.physik.uni-mainz.de}
\renewcommand{\thefootnote}{\dagger}
\footnotetext{e-mail: melic@thphys.irb.hr}
\renewcommand{\thefootnote}{\ddagger}
\footnotetext{e-mail: mereb@sun20.hepi.edu.ge}

\newpage

\section{Introduction}

Some time ago, after QCD started to become established as the theory of
hadronic interactions, 
a number of authors looked into the possibility of measuring $T$-odd effects
in current-induced 
interaction
\footnote{The phrase $T$-odd observables refers to observables that change
sign under simultaneous reflection of particle momenta and spins and
does not refer to truly $T$-violating observables.} 
that would result
from QCD rescattering effects. The rescattering effects were calculated 
from the absorptive parts of the relevant next-to-leading order 
QCD one-loop contributions.

The authors of \cite{RPL} considered $T$-odd effects in the decay of a
$J^{PC} = 1^{--}$ 
quarkonium state into three gluonic jets. $T$-odd effects in
$e^{+} e^{-}$-annihilation 
into three partonic jets were considered in \cite{KKSF,FSKS,BDS} 
excepting quark loop contributions. 
First, it came as a surprise that, for mass zero quarks, there are no
leading order ${\cal O}(\alpha_s^2)$
$T$-odd effects in this reaction \cite{KKSF}. This was understood more
systematically 
later on in \cite{HHK1,KS} from the observation that the absorptive parts
are necessarily 
proportional to the Born term 
and are thus unobservable. Measurable $T$-odd effects in these reactions
are generated by 
quark mass effects which were investigated in \cite{FSKS,BDS}. The
non-observability of 
${\cal O}(\alpha_s^2)$ $T$-odd effects in $e^{+} e^{-}$-interactions for
the massless case 
does not carry over to the crossed channels of deep inelastic scattering
(DIS) and the Drell-Yan (DY) process.
$T$-odd effects in DIS were explored in \cite{HHK1,HHK2} and in the 
DY process in \cite{HHK3}.

Since the early proposals to measure $T$-odd effects in current induced
interactions 
in the early eighties experimental facilities and techniques have
considerably been 
improved. Luminosities of lepton-hadron and hadron-hadron colliders have
dramatically 
increased providing for much higher event rates than was possible in the
earlier experiments. The energy range of the collliders has been extended
such that high momentum 
transfers can now be routinely probed. For example, at HERA one is starting
to probe weak 
interaction $Z$-exchange effects in neutral current events at very high
momentum transfers. 
This opens the door for the investigation of $T$- and $P$-odd effects in
neutral current DIS. 
Powerful jet finding and flavour tagging algorithms have been developed
that allow 
one to define asymmetry measures related to $T$-odd effects in DIS that
use parton jet 
observables instead of the semi-inclusive particle observables used in
the calculation of \cite{HHK1,HHK2,AG}.
Finally, there have been dramatic improvements in the availability of
polarized beams which 
again can be utilized to define new $T$-odd observables \cite{exper}.

It is therefore timely to take a fresh look at the subject of $T$-odd
observables in current-
induced reactions generated by QCD rescattering effects or, in a different 
language, by the absorptive parts of the corresponding one-loop
contributions. 
In this paper we point out that the ${\cal O}(\alpha_s^2)$ absorptive parts
of the 
relevant one-loop contributions in DIS and in the DY process can be
obtained through 
crossing from the well-known one-loop contributions to
$e^+ e^- \rightarrow q \overline q g$ 
annihilation calculated in \cite{KS}. This is theoretically appealing
and provides an 
independent check of the results presented before in DIS \cite{HHK1,HHK2}
and DY \cite{HHK3}. 
We also fill out some small odds and ends on the subject of $T$-odd
observables in these 
reactions which had not been covered in the earlier publications.

Our paper is structured as follows. In Sec. 2 we derive crossing rules
that allow one to cross from the 
$e^+ e^-$-annihilation channel to the DIS and DY channels. To obtain the
absorptive parts in the respective channels it is 
necessary to discuss the analyticity structure of the one-loop
contributions in the complex plane of the relevant kinematical variables.
The absorptive parts originate 
from logarithmic and dilogarithmic functions in the one-loop amplitude
when their arguments take
values on cuts 
in the analytic plane. We identify the range of values of the kinematical
variables in the three 
processes and show how to analytically continue the one-loop functions
from the $e^+e^-$-channel to the 
DIS and DY channels. Sec. 3 is devoted to a detailed discussion of $T$-odd
effects in DIS. 
We first provide a complete list of the nonsingular one-loop contributions
in $e^+e^-$-annihilation. 
Using the crossing rules laid down in Sec. 2 we analytically continue
the one-loop amplitudes to DIS. 
The absorptive one-loop amplitudes are then folded with the Born term
contributions. 
There are three $T$-odd hadronic structure functions,
$H_5$, $H_8$, and $H_9$, whose 
functional form is given for the quark, antiquark and gluon initiated cases.
We then define 
helicity structure functions which appear as angular coefficients in the
angular decay 
distribution of the DIS process when the hadronic tensor is folded with the
leptonic tensor. 
This allows us to compare our results with the results of \cite{HHK1,HHK2}.
We find complete 
agreement with the  $T$-odd results presented in these papers. In Sec. 4
we do the crossing 
and the analytical continuation to the DY process. 
Again we find agreement of our
results for the 
$T$-odd structures with those given in \cite{HHK3} 
after corrections for a 
typographical error reported in \cite{HKY1}.
In Sec.~5 we give our summary and provide an outlook on possible further
applications of our results to spin-dependent $T$-odd observables. In an
Appendix we present results on the dispersive and absorptive contributions
of the triangle anomaly graph to the DIS process.

\section{General principles of analyticity and crossing}

In this chapter we will develop the framework necessary to obtain the 
${\cal O}(\alpha_s^2)$ one-loop corrections in the DIS and DY
channels from the known results in the
$e^{+}e^{-}$-annihilation channel \cite{KS}.
In particular, we will derive crossing rules that allow one to determine
the whole
set of invariant hadronic structure functions $H_i$ (i=1,...,9), including
the absorptive $T$-odd structure functions for the DIS and the DY processes. 

In the absence of polarisation the definition of $T$-odd observables in
current-induced interactions
involves the analysis of parton processes with at least three partons
necessary to form triple momentum products. 
In order to fix our notation let us write down the momentum configuration
for $e^+ e^-$ annihilation into a quark, antiquark and a gluon:
\begin{equation}
\label{ll}
l^{-}(k)+l^{+}(k') \stackrel{\gamma,Z}{\longrightarrow} 
q(p_1)+\bar{q}(p_2)+g(p_3)
\end{equation}
where $l^{-}$ and $l^{+}$ are massless leptons, $q$ and $\bar{q}$ are
massless
quarks and antiquarks, respectively, and $g$ is a bremsstrahlung gluon. 
The momentum of the time-like gauge boson
is determined from four-momentum conservation and is given by $q=p_1+p_2+p_3$.

\begin{figure}
  \centerline{\epsfig{file=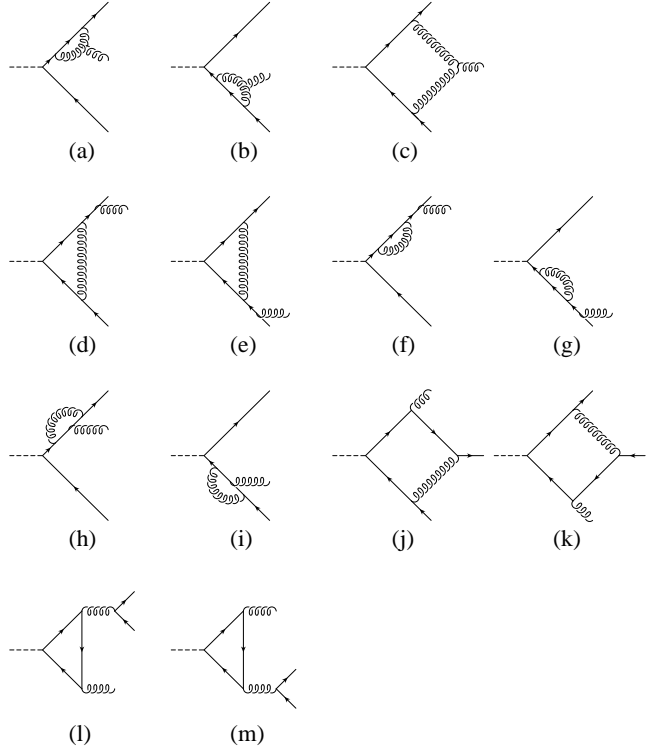,height=10cm,width=8.5cm,silent=}}
 \caption{${\cal O}(\alpha_s^2)$ one-loop corrections to the $e^+ e^-$
                                       annihilation.}
 \label{f:LOOP}
\end{figure}

The leading order contributions to the $T$-odd observables come from the 
interference of the absorptive parts of the one-loop amplitudes and the Born 
term amplitude. In Fig.~\ref{f:LOOP} we show 
the ${\cal O}(\alpha_s^2)$ one-loop diagrams that contribute to 
$e^+ e^- \rightarrow q \bar{q} g$ (with the leptonic part omitted).
They divide into the eleven 
contributions without quark loops and the two diagrams with a quark
loop. In the main part of this paper we will be mostly concerned with the
first eleven non-quark loop contributions (a) - (k). A discussion of
the so-called triangle-anomaly quark loop contributions (l) and (m)
is deferred to the Appendix.

At the three-parton level DIS proceeds through the following three 
subprocesses:
\begin{eqnarray}
\label{dis}
l(k)+q(p) & \stackrel{\gamma,Z,W^{\pm}}{\longrightarrow} &
l'(k')+q(p')+g(p_3),   \nonumber \\
l(k)+\bar{q}(p) & \stackrel{\gamma,Z,W^{\pm}}{\longrightarrow} & 
l'(k')+\bar{q}(p')+g(p_3),  \nonumber \\
l(k)+g(p)& \stackrel{\gamma,Z,W^{\pm}}{\longrightarrow} &
l'(k')+q(p')+\bar{q}(p_3) .
\end{eqnarray}
They are referred to as the quark-, antiquark- and gluon-initiated DIS 
processes, respectively. The momentum of the space-like gauge boson
is now given by $q=-p+p'+p_3$.

In the DY process the next-to-leading order contributions come from the 
following subprocesses:
 
\begin{eqnarray}
q(p_a)  + \overline{q}(p_b) & \stackrel{\gamma,Z,W^{\pm}}{\longrightarrow} &
l(k) + l^{\prime}(k') + g(p_3),  \label{dy1} \\
q(p_a)  + g(p_b) & \stackrel{\gamma,Z,W^{\pm}}{\longrightarrow} &
l(k) + l^{\prime}(k') + q(p_3),  \label{dy2} \\
\overline{q}(p_a)  + g(p_b) & \stackrel{\gamma,Z,W^{\pm}}{\longrightarrow} &
l(k) + l^{\prime}(k') + \overline{q}(p_3),  \label{dy3} 
\end{eqnarray}
where (\ref{dy1}) is the annihilation subprocess and (\ref{dy2}) 
and (\ref{dy3}) are the so-called quark- and antiquark-initiated "Compton" 
subprocesses, respectively. The momentum of the time-like gauge boson 
is given by $q=p_a+p_b-p_3$. Differing from $e^+e^-$-annihilation
there are also charged gauge boson contributions to DIS and the
DY process.

In what follows we need to discuss only the hadronic part of the three-parton
processes listed in (\ref{dis}) and (\ref{dy1}-\ref{dy3}). 
The contraction with the leptonic part will lead to angular factors 
and some $y$-dependence in the DIS case. The contraction with the
leptonic tensor will be discussed in the subsequent sections when we 
compare our results with the calculation of Hagiwara et al.
\cite{HHK1,HHK2,HHK3}. 

The relevant one-loop contributions to DIS and the DY process can be 
obtained from the one-loop contibutions to 
$e^+e^-$-annihilation calculated in \cite{KS} through crossing, i.e.
through the exchange of 
incoming and outgoing particle momenta in the one-loop diagrams in
Fig.\ref{f:LOOP}. 
For the real parts of the one-loop contributions and for the Born term 
contribution crossing can be implemented in a straightforward manner.
Crossing is more subtle for the imaginary parts of the one-loop amplitudes 
and needs a careful discussion of the analyticity properties of the 
one-loop amplitude.

The crossing of external lines in Feynman diagrams
implies
a sign change of the four momentum associated with that line. Thereby, the
values of the kinematic variables associated with the respective
momentum undergoes a discontinuous change. Massless one-loop amplitudes
contain
log and dilog functions  which depend on these kinematic
variables and which may be indefinite in certain ranges of their
domains of definition, i.e. they may be multivalued. One can choose
among the possible values by defining the value of the function at
a given point. Starting from this point one determines the value of the
multivalued function on the cuts by analytic continuation. The kinematic
variables are taken to be complex in this procedure.

In order to obtain a smooth
continuation one makes use of the imaginary parts of the one-loop
amplitudes given by the
($i\tau$)-form of the relevant propagators. 
In this way one avoids possible ambiguities. As an
example we take the natural logarithm. 
The logarithm is taken as a
complex-valued analytic function with a cut on the negative real axis
and a branch point located at zero.

Next consider the natural logarithm of an arbitrary positive real number
$x$. To obtain its value at ($-x$) we use

\[  \ln(-x)=\ln(|x|)+i(2k+1)\pi,   {\rm \hspace{.2in}}  {\rm with} 
              {\rm \hspace{.2in}}     k \in {\cal Z}. \]

The integer number $k$ will be determined from the phase angle of the
complex number ($-x$). If one excludes multiple
rotations in the complex plane then one remains with only two
possibilities: $k=0,-1$. For all complex numbers of the form
($-x\pm i\tau$) with an infinitesimally small positive $\tau$ one would
have the following identity:
\begin{equation}
\label{log}
   \lim_{\tau \rightarrow 0} \ln(-x\pm i\tau)=\ln(|x|)\pm i\pi .
\end{equation}

The results of the calculation of the one-loop contributions to
$e^+ e^-$-annihilation listed in
\cite{KS} contain no explicit $i\tau$ prescription. This is adequate
for the $e^+ e^-$-reaction since in this case the results are given for
regions in
the complex plane away from the singularities. 
The results are valid only in this restricted region and need to be  
analytically
continued to the other regions in the complex plane 
accessible in DIS and in the DY case. 
It is, however, 
possible to restore the omitted $i\tau$ prescriptions in \cite{KS}
in a straightforward way.
For the relevant kinematic variables the infinitesimal imaginary
parts are provided by the Feynman rules if one takes the full propagators 
in their original ($i\tau$)-form. In this case one has the following 
terms from solving the loop integrals 
$(s_{ij}\equiv (p_i+p_j)^2=2 p_ip_j)$:
\begin{equation}
\frac{s_{ij}+i\tau}{q^2+i\tau}, {\rm \hspace{.2in}}
1-\frac{s_{ij}+i\tau}{q^2+i\tau}, {\rm \hspace{.2in}}
{\rm and} {\rm \hspace{.2in}}   (-q^2-i\tau)^{-\varepsilon}, 
\end{equation}
where ${\varepsilon}=2-d/2$, and $d$ is the continuous space-time
dimension.

Here one should notice that the form of energy-momentum conservation for
the s-channel annihilation ensures relative plus signs between the
three scalar invariants
$s_{ij}=2p_ip_j$ and $i\tau$, as well as between $q^2$ and $i\tau$ in 
the denominators of the respective
Feynman integrals. 
Thus, more generally, one has the following rules for
s-channel annihilation:
\begin{equation}
\label{general}
q^2 \rightarrow q^2+i\tau, {\rm \hspace{.4in}}
s_{ij} \rightarrow s_{ij}+i\tau .
\end{equation}                       
With these rules the results of \cite{KS} are valid in any kinematical
region.

For the kinematical variables $y_{ij}=s_{ij}/q^2$ 
used in \cite{KS} one finds the following replacements:
\begin{eqnarray}
\label{rules}
\nonumber
y_{ij}&\rightarrow&y_{ij} + \frac{i\tau}{q^2}(1-y_{ij}) +
                                     {\cal O}(\tau^2),   \\
1-y_{ij}&\rightarrow&1-y_{ij} - \frac{i\tau}{q^2}(1-y_{ij}) + {\cal
O}(\tau^2),  \\
\nonumber
(-q^2)^{-\varepsilon}&\rightarrow&1-\varepsilon \ln(-q^2-i\tau) +
                         {\cal O}(\varepsilon^2).
\end{eqnarray}

For every contributing subprocess in DIS and DY one has to perform a detailed
investigation of the range of values of the $y_{ij}$'s after crossing and
then one can analytically continue the log and dilog functions
and thereby remove the ambiguity
which occurs when one changes the sign of their arguments.
The analytic continuation of the logarithm function is given in (\ref{log}). 
For the dilogarithms, when $x>1$, we use the
identity
\begin{equation}
\label{dilog}
{\rm Li}_2(x)=-{\rm Li}_2(1-x)+\zeta(2)-\ln(x)\ln(1-x),  
\end{equation}
and treat the complex logarithm $\ln(1-x)$ according to (\ref{log}).

At this point we introduce the usual hadronic DIS variables $x$ and $z$
\begin{equation}
\label{vardis}
x=\frac{-q^2}{2qp}, {\rm \hspace{.45in}}  z=\frac{pp'}{qp} \, ,
\end{equation} 
and proceed with the crossing procedure as described above. 
The crossing from the $e^+ e^-$-channel to the quark-initiated
subprocesses 
in DIS is given by the following change of the momenta 
\begin{equation}
\label{mDISq}
p_{1} \rightarrow p^{\prime} \qquad {\rm and} \qquad
p_{2} \rightarrow -p \, ,
\end{equation}
according to the momentum definitions in (\ref{ll}) and (\ref{dis}). 
The change of the kinematical variables $y_{ij}$'s and the new ranges of 
their values are given by
\begin{eqnarray}
\label{qini}
\nonumber
     y_{12} \equiv \frac{2 p_1 p_2}{q^2}\rightarrow y_{12}^{q} 
      \equiv \frac{2 p p'}{-q^2} 
&=& \frac{z}{x} {\rm\hspace{.45in}}   
                                  \in [1,\infty] \, , \\
     y_{13} \equiv \frac{2 p_1 p_3}{q^2}\rightarrow y_{13}^{q} 
      \equiv \frac{2 p' p_3}{q^2}
&=& 1-\frac{1}{x} {\rm\hspace{.2in}} 
                                  \in [-\infty,0] \, ,\\
\nonumber
     y_{23} \equiv \frac{2 p_2 p_3}{q^2}\rightarrow y_{23}^{q} 
      \equiv \frac{2 p p_3}{-q^2}
&=& \frac{1-z}{x} {\rm\hspace{.2in}}
                                  \in [1,\infty] \, .
\end{eqnarray}
One should note that $q^2 < 0$ in the crossed DIS channel. The 
corresponding crossing to the antiquark-initiated subprocess involves the
momentum changes $p_1 \rightarrow -p$ and $p_2 \rightarrow p'$.
However, we need not explicitly discuss crossing for this case since one
can use $CP$-invariance in the final result for the quark-initiated case
to obtain the corresponding antiquark-initiated results.  

Similarly, the crossing from the $e^+e^-$-annihilation to the gluon-initiated
subprocess in DIS is effected by 
\begin{equation}
\label{mDISg}
p_{1} \rightarrow p^{\prime} \qquad {\rm and} \qquad 
p_{3} \rightarrow -p \, .
\end{equation}
The resulting $y_{ij}^g$'s are:
\begin{eqnarray}
\label{gini}
\nonumber
     y_{12}\rightarrow y_{12}^{g} \equiv \frac{2 p' p_2}{q^2}
&=& 1-\frac{1}{x} {\rm\hspace{.2in}}
                                  \in [-\infty,0], \\
     y_{13}\rightarrow y_{13}^{g} \equiv \frac{2 p p'}{-q^2}
&=& \frac{z}{x} {\rm \hspace{.45in}}
                                  \in [1,\infty], \\
\nonumber
     y_{23}\rightarrow y_{23}^{g} \equiv \frac{2 p p_2}{-q^2}
&=& \frac{1-z}{x} {\rm\hspace{.2in}}
                                  \in [1,\infty].
\end{eqnarray}

The one-loop results in \cite{KS} are presented in terms of the
variables $x_{k}=2 p_k q/q^2$ $(k=1,2,3)$. The $x_k$ are related to the
$y_{ij}$ via 
\[       x_{k}=1-y_{ij},   {\rm\hspace{.4in}}   (k \not= i,j).      \]

Next we turn to the Drell-Yan process. For the annihilation
subprocess (\ref{dy1}) crossing implies the following 
change of momenta
\begin{equation}
\label{mDYa}
p_{1} \rightarrow -p_{a} \qquad {\rm and} \qquad 
p_{2} \rightarrow -p_{b} \, .
\end{equation}
For the quark-initiated "Compton" scattering one has to change
\begin{equation}
\label{mDYC}
p_{1} \rightarrow -p_{a} \qquad {\rm and} \qquad 
p_{3} \rightarrow -p_{b} \, ,
\end{equation}
with
$q \rightarrow -q$ in both cases. Again we omit explicit
reference to the antiquark-initiated "Compton" scattering case because
its structure follows from the quark-initiated "Compton"
scattering case through $CP$ invariance.

The corresponding kinematical variables in the annihilation subprocesses are now
\begin{eqnarray}
\label{ann}
\nonumber
     y_{12}\rightarrow y_{12}^{a} \equiv \frac{2 p_a p_b}{q^2}
&=& \frac{1}{x_a} + \frac{1}{x_b} -1 {\rm\hspace{.2in}}
                                  \in [1,\infty] \, ,\\
     y_{13}\rightarrow y_{13}^{a} \equiv -\frac{2 p_a p_3}{q^2}
&=& 1 - \frac{1}{x_b} {\rm \hspace{.50in}}
                                  \in [-\infty , 0] \, ,\\
\nonumber
     y_{23}\rightarrow y_{23}^{a} \equiv -\frac{2 p_b p_3}{q^2}
&=& 1-\frac{1}{x_a} {\rm\hspace{.5in}}
                                  \in [-\infty , 0] \, .
\end{eqnarray}
The crossed $y_{ij}$'s for the quark-initiated "Compton" subprocesses are
\begin{eqnarray}
\label{Com}
\nonumber
     y_{12}\rightarrow y_{12}^{C_q} \equiv -\frac{2 p_a p_3}{q^2}
&=& 1-\frac{1}{x_b} {\rm\hspace{.6in}}
                                  \in [-\infty,0] \, ,\\
     y_{13}\rightarrow y_{13}^{C_q} \equiv \frac{2 p_a p_b}{q^2}
&=& \frac{1}{x_a} + \frac{1}{x_b} -1 {\rm \hspace{.35in}}
                                  \in [1,\infty] \, , \\
\nonumber
     y_{23}\rightarrow y_{23}^{C_q} \equiv -\frac{2 p_b p_3}{q^2}
&=& 1-\frac{1}{x_a} {\rm\hspace{.6in}}
                                  \in [-\infty , 0] \,.
\end{eqnarray}

\narrowtext

\begin{table}
\caption{Imaginary parts of different functions appearing after crossing from 
$e^+ e^-$-annihilation to the quark- and gluon-initiated cases in DIS,
denoted by $\rm DIS^q$ and 
$\rm DIS^g$, respectively, and to 
the annihilation ($\rm DY^a$) and "Compton" ($\rm DY^C$) subprocesses in
DY. The invariants $y_{ij}^q$, $y_{ij}^g$, $y_{ij}^a$ and $y_{ij}^C$ are
defined in (\ref{qini}), (\ref{gini}), (\ref{ann}) and (\ref{Com})}

\begin{tabular}{cccccc}
functions & $\rm DIS^q$  & $\rm DIS^g$ & &  $\rm DY^a$  & $\rm DY^C$\\ 
\hline
$\ln(y_{12})$  &  0 & $-i \pi$ & & 0 &  $i \pi$ \\
$\ln(1 - y_{12})$ & $-i \pi$ & 0 & & $i \pi$ & 0 \\
${\rm Li}_2(y_{12})$ & $i \pi \ln(y_{12}^q)$ & 0 & & $-i \pi
\ln(y_{12}^a)$ & 0\\
\hline
$\ln(y_{13})$  &  $-i \pi$  & 0 & & $i \pi$ & 0\\
$\ln(1 - y_{13})$ & 0 &  $-i \pi$  & & 0 & $i \pi$\\
${\rm Li}_2(y_{13})$ & 0 & $i \pi \ln(y_{13}^g)$ & & 0 & $-i \pi
\ln(y_{13}^C)$ \\
\hline
$\ln(y_{23})$  &  0 & 0 & &  $i \pi$ & $i \pi$\\
$\ln(1 - y_{23})$ & $-i \pi$ & $ -i \pi$  & & 0 & 0  \\
${\rm Li}_2(y_{23})$ & $i \pi \ln(y_{23}^q)$ & $i \pi \ln(y_{23}^g)$ & & 0
& 0
\end{tabular}
\label{t:tab1}
\end{table}

In the above equation we have introduced 
the DY variables $x_a$ and $x_b$  defined by
\begin{equation}
\label{vardy} 
x_a=\frac{M^2}{2qp_a}, {\rm \hspace{.45in}}  x_b=\frac{M^2}{2 qp_b} \, ,
\end{equation} 
where $M^2$ is the mass of the exchanged gauge bosons $W^{\pm}$ and $Z$,
or, 
for the electromagnetic interaction, one has $M^2 = q^2$.

\begin{figure}
  \centerline{\epsfig{file=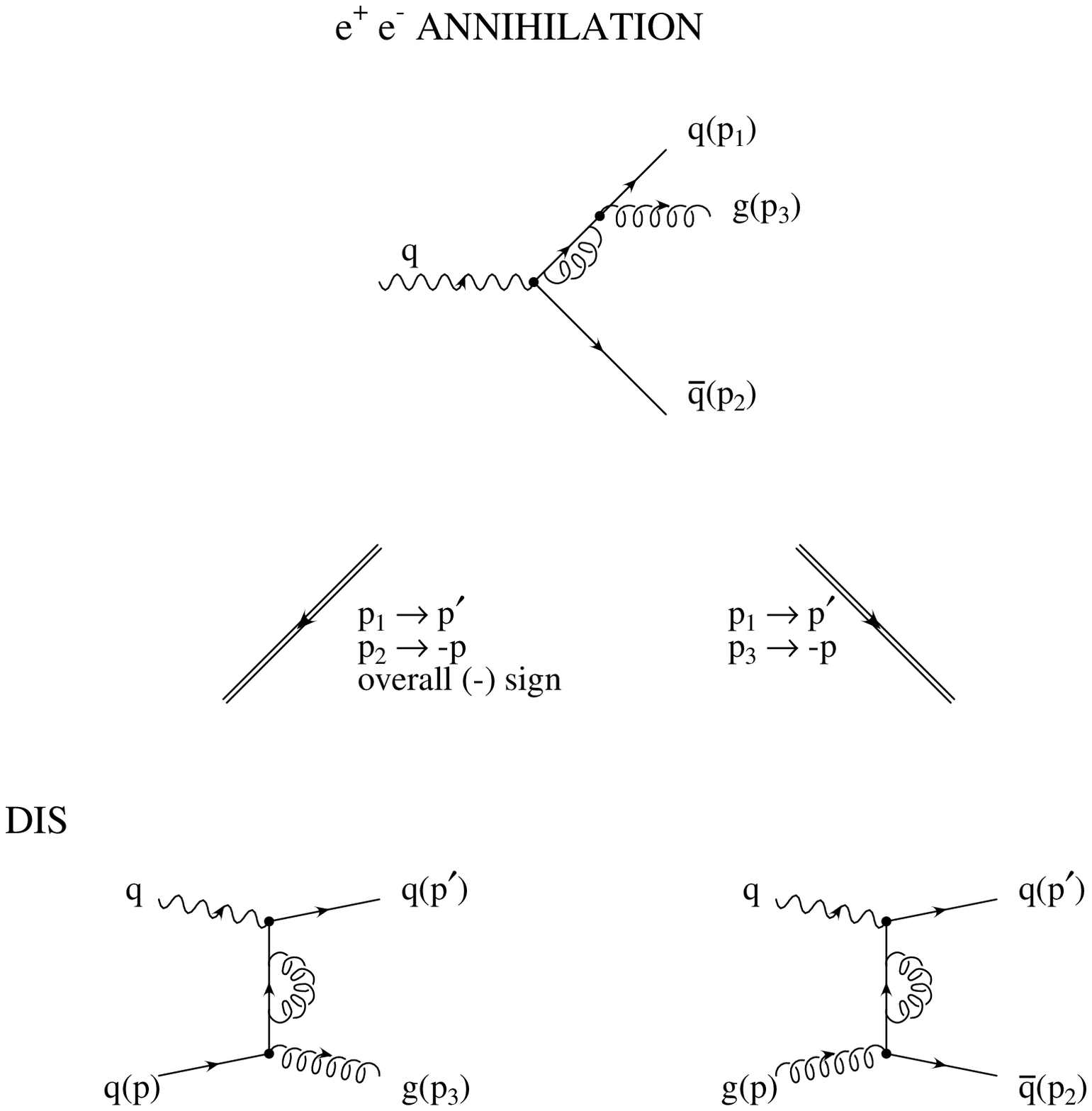,height=9cm,width=8.5cm,silent=}
              \epsfig{file=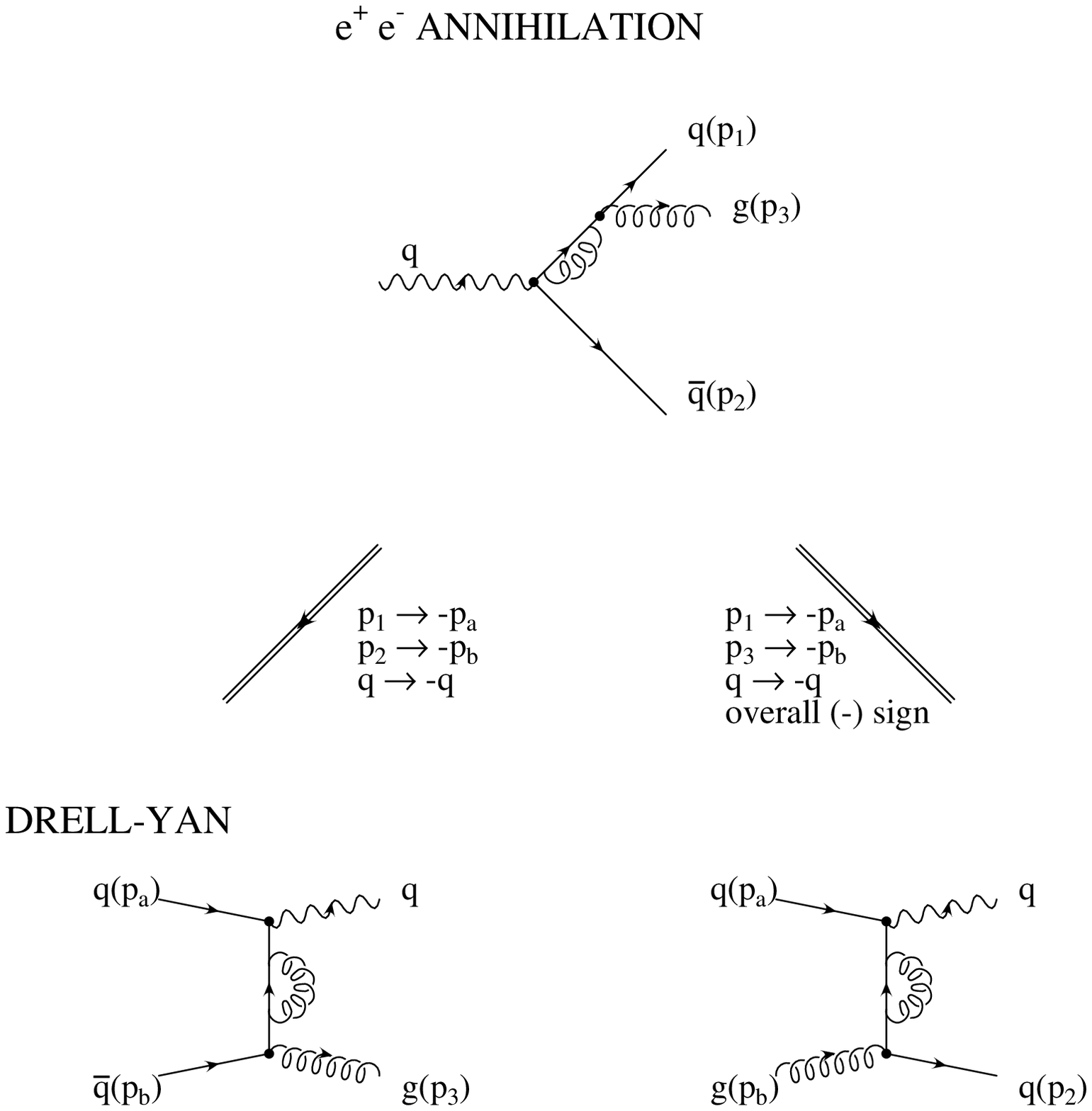,height=9cm,width=8.0cm,silent=}}
 \caption{ Illustration of crossing from 
           $e^+ e^-$-annihilation to the DIS and DY processes
           for one representative one-loop contribution. Shown are
           crossing to the quark- and gluon-initiated DIS processes
           and the annihilation and quark-initiated Compton processes
           in DY.}
 \label{f:cross}
\end{figure}

Note the $2 \leftrightarrow 3$ symmetry between the annihilation and the
quark-initiated Compton subprocess in terms of the 
subprocess variables $x_a$ and $x_b$.
The momentum changes involved in crossing from the $e^+ e^-$-channel to
DIS and the DY channel are summarized in Fig. \ref{f:cross} using diagram
1(f) as an illustrative example.

As we have already mentioned before, the results of the calculation
in $e^+ e^-$-annihilation 
channel contain log and 
dilog functions, which, after applying the rules from
(\ref{qini}-\ref{gini}) and 
from (\ref{ann}-\ref{Com}) have to be analytically continued to the
new kinematically 
allowed regions. Using the cut structure defined in (\ref{log}) and
(\ref{dilog}), one obtains their absorptive 
parts. The results of crossing and analytic continuation of logarithmic and 
dilogarithmic functions appearing in our calculation of $T$-odd amplitudes 
in DIS and DY channels are summarized in Table \ref{t:tab1}.

\section{Crossing Results for Hadronic Structure Functions in DIS}

We now proceed to derive the $T$-odd hadronic structure functions in DIS 
by applying the analyticity and crossing rules derived in the proceeding
section starting with the $e^+ e^-$ results in \cite{KS}. We will be
mostly
interested in the imaginary (absorptive) parts but will also briefly
comment on the crossing properties of the real parts of the one-loop
amplitudes and the Born term amplitudes. The results of the 
crossing procedure will then be compared to
the corresponding results in DIS in the three quark, antiquark 
and gluon-initiated cases \cite{HHK1,HHK2}. 
As was mentioned in the introduction, the $T$-odd structure
functions in the $e^+e^-$ case vanish identically at
the one-loop level for the set of graphs (1a)-(1k) with no quark 
loops\footnote{There are contributions to the $T$-odd structure functions
coming from the quark loop graphs (l) and (m) in Fig.~\ref{f:LOOP} even
for massless external
quarks due to an incomplete cancellation of the $b$- and $t$-quark in
the quark loop. These contributions have been shown to be very small
\cite{HKY2}}. 
The reason is that the absorptive parts of the $e^+ e^-$-annihilation
massless one-loop amplitudes are proportional to the singular terms of its
dispersive part which in turn have Born term structure \cite{KKSF,KS}. The
absorptive parts are thus not measurable at this order of perturbation
theory.
However, the kinematics is different in the crossed processes and this
proportionality no longer holds leading to nonvanishing $T$-odd effects
in the crossed channels.

We begin our discussion by recapitulating the $e^+ e^-$-annihilation
one-loop results given in \cite{KS}. They are needed
as a starting point for the crossing procedure.
The transition amplitude $T^V_{\mu\beta}$ for the vector current
transition
$V_{\mu} \rightarrow q\bar{q} g$ can be expanded in
$d$-dimensions along the seven independent covariants \cite{KS} 
\begin{eqnarray}
\label{exp}
\nonumber
T^V_{\mu\beta}&=&N_i \; C^i_{\mu\beta} \qquad\qquad\qquad (i=1,...,7)    \\
\nonumber
&=& q^{-4} N_1 \;\hat{p}_{+\mu} \tilde{p}_{-\beta} \not p_3 +
    q^{-4} N_2 \;\hat{p}_{-\mu} \tilde{p}_{-\beta} \not p_3  \\
\nonumber
& & + q^{-2} N_3 \;\hat{\gamma}_{\mu} \tilde{p}_{-\beta} +
      q^{-2} N_4 \;\hat{p}_{+\mu} \tilde{\gamma}_{\beta}  \\
\nonumber
& & + q^{-2} N_5 \;\hat{p}_{-\mu} \tilde{\gamma}_{\beta}  +  
      q^{-2} N_6 \;\hat{ \tilde{g} }_{\mu\beta} \not p_3   \\
& & + N_7 \left( \gamma_{\mu} \frac{\not p_2+\not p_3}{s_{23}}\gamma_{\beta}
   - \gamma_{\beta} \frac{\not p_1+\not p_3}{s_{13}} \gamma_{\mu} \right)
\end{eqnarray}
where $q=p_1+p_2+p_3$ and $p_{\pm}=p_1 \pm p_2$.
The covariants $C^i_{\mu\beta}$ are defined through Eq.~(\ref{exp}) and 
the symbols ``$\sim$'' and ``$\wedge$'' denote the gauge
invariant completions 
\begin{eqnarray}
\nonumber
\hat{p}_{\mu}&=&p_{\mu}-\frac{pq}{q^2} q_{\mu}  \, ,       \\
\nonumber
\hat{\gamma}_{\mu}&=&\gamma_{\mu}-\frac{\not q}{q^2} q_{\mu} \, , \\
\nonumber
\tilde{p}_{\beta}&=&p_{\beta}-\frac{pp_3}{p_3q} q_{\beta} \, ,   \\
\nonumber
\tilde{\gamma}_{\beta}&=&\gamma_{\beta}-\frac{\not p_3}{p_3q} q_{\beta} \, , \\
\hat{ \tilde{g} }_{\mu\beta}&=&g_{\mu\beta}-\frac{q_{\beta}p_{3\mu}}{p_3q} \, .
\end{eqnarray}
In writing down the covariant expansion it is understood that the
covariants are taken between the relevant spin wave functions, i.e.
$\bar{u}_1 T^V_{\mu\beta} v_2 \varepsilon^{*\beta}$.

We emphasize that the $T$-odd structure functions resulting from the
one-loop amplitudes are infrared (IR) and collinear (M) finite in $d=4$
dimensions. This
follows indirectly from the Lee-Nauenberg theorem in that there are no
corresponding tree graph contributions that could cancel the divergencies
if there were any. We could in principle therefore keep $d=4$ in
our calculation. In this case one has overcounted the number of covariants
in the above expansion. There are in fact only six independent covariants
in four dimensions as can be verified by counting the number of
independent helicity amplitudes. For the sake of completeness we list
the linear relation between the seven
covariants $C^i_{\mu\beta}$ in $d=4$ dimensions (taken between spin
wave functions) which can be obtained
from\footnote{We take this opportunity to correct Eq. (A.7) in \cite{KSi}.
Eq. (A.7) in \cite{KSi} should read:\\ 
${\rm \hspace{.2in}} 
(1-x_3)k^7_{\beta\mu}=k^1_{\beta\mu}-\frac{1}{2}x_3(k^3_{\beta\mu}-
k^5_{\beta\mu})-\frac{1}{2}(2x_1-x_3)k^4_{\beta\mu}+(1-x_3)k^6_{\beta\mu}$.}
\cite{Per,KSi}. One has
\begin{eqnarray}
(1-x_1)(1-x_2)(1-x_3)C^7_{\mu\beta} &=& x_3 C^1_{\mu\beta} -
\frac{1}{2} x_3 (x_1 + x_2) C^3_{\mu\beta} 
- \frac{1}{2} ( x_1^2 - x_2^2)  C^4_{\mu\beta}  \nonumber\\
& & +
\frac{1}{2} x_3^2  C^5_{\mu\beta} -
(x_1 -x_2)(1-x_3)  C^6_{\mu\beta}.
\end{eqnarray}
As concerns the present application it is 
nevertheless technically advantageous to work with the (overcounted)
set of the above seven covariants. Note that the seventh covariant
$C^7_{\mu \beta}$ in (\ref{exp})
has been chosen to have Born term structure. This will be important to
keep in mind in what follows.   
As we are dealing with massless quarks the case of axial vector current
transition $A_{\mu} \rightarrow q\bar{q} g$ can be easily dealt with. We
simply have to multiply the vector current amplitude by $\gamma_{5}$
from the left. One has
\begin{equation}
\label{aexp}
T^A_{\mu\beta}=\gamma_{5} T^V_{\mu\beta} \, .
\end{equation}

What is needed are explicit forms of the one-loop amplitudes
in the $e^+e^-$ channel. The IR and M singular one-loop contributions have
Born term structure and will not be needed in the following. As concerns
the nonsingular one-loop contributions, 
we decided to reproduce the two relevant 
tables, Table~\ref{t:tab2} and Table~\ref{t:tab3}, from \cite{KS} because
the results of \cite{KS} may not be readily available to everyone (in
those early days there was no electronic publishing). 

There are two colour structures in the
one-loop amplitudes referred to as 
the QCD and QED structure. The corresponding amplitudes are denoted by
$\tilde{N}_i$ (Table~\ref{t:tab2}) and $\hat{N}_i$ (Table~\ref{t:tab3}),
respectively,
\[    N_l^{(j)}=g^3 \{ \tilde{N}_l i f^{ijk} t_k t_i +
                       \hat{N}_l t_i t_j t_i \}.            \]

\begin{table}[ht]
\squeezetable
\caption{ Nonsingular ''QCD'' contributions to seven invariant amplitudes
$\tilde N_i$. 
Entries denoted by $\leftarrow s$ are obtained from their left neighbours 
by interchanging 1 and 2. Correspondingly for the entries $\leftarrow a$
with an additional minus sign. 
$\displaystyle x_i = \frac{2 p_i q}{q^2}$, $i = 1,2,3$ and 
$R(x,y) = \displaystyle \ln(x) \ln(y)- \ln(x) \ln(1-x) - 
\ln(y) \ln(1-y) - {\rm Li}_2(x) - {\rm Li}_2(y) + \pi^2/6$, where 
$\displaystyle {\rm Li}_2(x) = - \int_0^x dz \frac{\ln(1-z)}{z}$ . }

\begin{tabular}{ccccc}
$\displaystyle QCD$ & $\displaystyle \frac{C_1}{2}$ & 
$ \displaystyle\frac{C_1}{2}\ln(y_{13})$ &
$\displaystyle\frac{C_1}{2}\ln(y_{23})$ & 
$\displaystyle\frac{C_1}{2}R(y_{13}, y_{23})$ 
\vspace{0.2cm} \\
\hline \\
$\tilde N_1$  & $\displaystyle\frac{x_1(1-x_2) + x_2 (1-x_1)}{x_1 x_2
(1-x_1)(1-x_2)}$ &
$\displaystyle\frac{(1-x_3) + 3 x_2 (-1 + x_1 - x_2)}{(1-x_1)(1-x_3)
x_2^2}$ & 
$\leftarrow s $ & $-\displaystyle\frac{6}{ (1-x_3)^2}$ \vspace{0.3cm}\\
\hline \\
$\tilde N_2$  & $-\displaystyle\frac{x_1-x_2}{x_1 x_2 (1-x_1)(1-x_2)}$ &
$\displaystyle\frac{ -3 x_2 -1 }{(1-x_1) x_2^2}$ & 
$\leftarrow a$ &  0  \vspace{0.3cm}\\
\hline \\
$\tilde N_3$  & $-\displaystyle\frac{x_3}{(1-x_1)(1- x_2)}$ &
$\displaystyle\frac{ 3 }{(1-x_3) x_2}$ & 
$\leftarrow s$ &  $\displaystyle\frac{3 (x_1+x_2)}{ (1-x_3)^2}$
\vspace{0.3cm}\\
\hline \\
$\tilde N_4$  & $\displaystyle\frac{x_1-x_2}{x_1 x_2}$ &
$\displaystyle\frac{2 x_2 (1+ x_1 - 2 x_2) - x_3 (1-x_1) }{x_2^2 x_3
(1-x_3)}$ & 
$\leftarrow a$ &  $\displaystyle\frac{3 (x_1^2-x_2^2)}{x_3 (1-x_3)^2}$
\vspace{0.3cm}\\
\hline \\
$\tilde N_5$  & $-\displaystyle\frac{x_1+x_2}{x_1 x_2}$ &
$\displaystyle\frac{1- x_1 - 4 x_2 }{x_2^2 (1-x_3)}$ & 
$\leftarrow s$ &  $-\displaystyle\frac{3 x_3}{ (1-x_3)^2}$  \vspace{0.3cm}\\
\hline \\
$\tilde N_6$  & 0  &
$\displaystyle\frac{6 (1-x_2) }{x_2 x_3}$ & 
$\leftarrow a$ &  $\displaystyle\frac{6 (x_1 -x_2)}{ x_3 (1-x_3)}$
\vspace{0.3cm}\\
\hline \\
$\tilde N_7$  &  7 & 
$\displaystyle \ln(y_{13}) + \frac{3 (1-x_1)(1-x_2)}{x_2x_3}$ &
$\leftarrow s$ &  $2 \displaystyle\frac{x_3 (1-x_3) + 3 (1-x_1)(1-x_2)}{
x_3 (1-x_3)}$ \\

\end{tabular}

\label{t:tab2}
\end{table}
 
As a next step one folds the above one-loop amplitude with the Born term
amplitude proportional to $C_{\mu\beta}^7$ and sums over the spins and
colors of the final partons.
The result can be expanded in terms of nine gauge invariant covariants
which define the nine invariant structure functions $H_i (i=1,...,9)$ of
the $e^+ e^-$-annihilation process.
\begin{eqnarray}
\label{hmunu}
\nonumber
H_{\mu \nu}&=& H_1 \left(g_{\mu \nu}-\frac{q_{\mu}
q_{\nu}}{q^2}\right) \\
\nonumber
& & + H_2 q^{-2} \hat{p}_{1\mu} \hat{p}_{1\nu}           \\
\nonumber
& & + H_3 q^{-2} \hat{p}_{2\mu} \hat{p}_{2\nu}           \\
\nonumber
& & + H_4 q^{-2} (\hat{p}_{1\mu} \hat{p}_{2\nu} + \hat{p}_{1\nu}\hat{p}_{2\mu})
\\
\nonumber
& & + H_5 q^{-2} (\hat{p}_{1\mu} \hat{p}_{2\nu} - \hat{p}_{1\nu}\hat{p}_{2\mu})
\\
& & + H_6 q^{-2} i \epsilon_{\mu\nu\alpha\beta} q^{\alpha} p_1^{\beta}    \\
\nonumber
& & + H_7 q^{-2} i \epsilon_{\mu\nu\alpha\beta} q^{\alpha} p_2^{\beta}    \\
\nonumber
& & + H_8 q^{-4} (\hat{p}_{1\mu} F_{\nu} + \hat{p}_{1\nu} F_{\mu})     \\
\nonumber
& & + H_9 q^{-4} (\hat{p}_{2\mu} F_{\nu} + \hat{p}_{2\nu} F_{\mu}),
\end{eqnarray}
where $
F_{\mu}=i \epsilon_{\mu\alpha\beta\gamma} p_1^{\alpha}p_2^{\beta}q^{\gamma}
$. Throughout this paper we use the convention $\epsilon^{0123}=-1$.

From the hermiticity property $H_{\mu \nu}=H_{\nu \mu}^*$ one concludes
that $H_1,H_2,H_3,H_4,H_6$ and $H_7$ are purely 
real and $H_5, H_8$ and $H_9$ are purely imaginary functions. Also,
the first five structure functions $H_1 - H_5$
are parity conserving and $H_6 - H_9$ are parity
violating. As was mentioned before the $T$-odd invariant structures $H_5, H_8$ 
and $H_9$ vanish in $e^+ e^-$-annihilation at this order of perturbation
theory for massless quarks.

\begin{table}
\squeezetable

\narrowtext
\caption{ Nonsingular ''QED'' contributions to 7 invariant amplitudes
$\hat N_i$. 
Here $X = (1-x_1)(1-x_2)(1-x_3)$ and all other notations as in the Table
\protect\ref{t:tab2}.} 

\begin{tabular}{ccccccc}

$\textstyle QED$ & $\textstyle \frac{C_1}{2}$ & $
\textstyle\frac{C_1}{2}\ln(y_{12})$ & 
$\textstyle\frac{C_1}{2}\ln(y_{13})$ &
$\textstyle\frac{C_1}{2}\ln(y_{23})$ &
$\textstyle\frac{C_1}{2}R(y_{12}, y_{13})$ &
$\textstyle\frac{C_1}{2}R(y_{12}, y_{23})$
\vspace{0.2cm} \\
\hline \\
$\textstyle \hat N_1$  & $\textstyle -\frac{x_1 + x_2 + 2 x_1 x_2 }{x_1
x_2 (1-x_1)(1-x_2)}$ &
$ \textstyle \frac{ \begin{array}{l} \scriptstyle -2 \{x_3 (1 + x_1 + x_2 \\ 
 \scriptstyle \qquad - x_1 x_2 -x_1^2 - x_2^2) - 2 X \}\end{array} }{x_3
(1-x_1)^2(1-x_2)^2}$ & 
$\textstyle\frac{\begin{array}{l} \scriptstyle \{-2(1-x_3)^2(x_2+ 3-2x_1)\\
\scriptstyle \qquad\qquad - 3 X + 4 (1-x_1)^3 \} \end{array} }{x_2^2
(1-x_1)^2(1-x_3) }$ & 
$\textstyle \leftarrow s $ & $\textstyle -2\frac{x_1+x_2}{ (1-x_1)^3}$ &
$\textstyle \leftarrow s $ \vspace{0.3cm}\\
\hline \\
$\textstyle \hat N_2$  & $\textstyle\frac{x_1-x_2}{x_1 x_2 (1-x_1)(1-x_2)}$ &
$\textstyle 2\frac{x_3(x_1-x_2)}{(1-x_1)^2(1-x_2)^2}$ & 
$\textstyle\frac{(1-x_1)(1+x_2)+ 2 x_2 x_3}{x_2^2(1-x_1)^2}$ & 
$\textstyle\leftarrow a$ & $\textstyle 2\frac{1-x_1+x_3}{(1-x_1)^3}$ &
$\textstyle\leftarrow a$ \vspace{0.3cm}\\
\hline \\
$\textstyle\hat N_3$  & $-\textstyle\frac{x_3}{(1-x_1)(1- x_2)}$ &
$\textstyle\frac{x_3}{(1-x_1)(1-x_2)}$ & 
$\textstyle\frac{x_2 x_3 - 3 (1-x_1)}{x_2(1-x_1)(1-x_3)}$ & 
$\textstyle\leftarrow s$ &  $\textstyle\frac{2 - 2 x_1 + x_3}{ (1-x_1)^2}$ &  
$\textstyle\leftarrow s$ \vspace{0.3cm}\\
\hline \\
$\textstyle \hat N_4$  & $\textstyle -\frac{x_1-x_2}{x_1 x_2}$ &
$\textstyle -\frac{x_1^2-x_2^2}{x_3(1-x_1)(1-x_2)}$ & 
$\textstyle\frac{ \begin{array}{l} \scriptstyle \{ 2 x_2^3(3-x_3) -
x_2^2(x_3^2 -11 x_3 +12) \\
\scriptstyle \qquad\qquad + (1-x_3)^2(6 x_2 + x_3) \} \end{array}}{x_2^2
x_3 (1-x_1)(1-x_3)}$ & 
$\textstyle\leftarrow a$ &  $\textstyle\frac{\begin{array}{l}
\scriptstyle \{-1 + x_2^2 \\
\scriptstyle \; -(1-x_1)(3-3 x_1 + 2 x_2) \} \end{array}}{x_3 (1-x_1)^2}$  &
$\textstyle\leftarrow a$ \vspace{0.3cm}\\
\hline \\
$\textstyle\hat N_5$  & $\textstyle\frac{x_1+x_2}{x_1 x_2}$ &
$\textstyle\frac{x_3}{(1-x_1)(1-x_2)}$ & 
$\textstyle\frac{x_2 (1-x_1+x_2 x_3) +(1-x_1)(1-x_3)}{x_2^2(1-x_1)(1-x_3)}$ & 
$\textstyle\leftarrow s$ &  $\textstyle \frac{x_3 + 2 (1-x_1)}{ (1-x_1)^2}$ &
$\textstyle\leftarrow s$  \vspace{0.3cm}\\
\hline \\
$\textstyle\hat N_6$  & $\textstyle 0$  &
$\textstyle 2\frac{(1-x_3)(x_1-x_2)}{x_3(1-x_1)(1-x_2)}$ & 
$\textstyle 2\frac{x_2 x_3 - 3 (1-x_1)(1-x_2)}{x_2 x_3(1-x_1)}$ & 
$\textstyle \leftarrow a$ &  $\textstyle  2\frac{x_3(1-x_3)- 2(1
-x_1)^2}{ x_3 (1-x_1)^2}$ & 
$\textstyle\leftarrow a$  \vspace{0.3cm}\\
\hline \\
$\textstyle\hat N_7$  & $\textstyle -9 $ & 
$\textstyle -\ln(y_{12})$ & $\textstyle -\frac{3 (1-x_1)(1-x_2)}{ x_2 x_3}$ &
$\textstyle\leftarrow s$ &  $\textstyle -2\frac{(1-x_1)}{ x_1 }$  &
$\textstyle \leftarrow s$ \\

\end{tabular}

\label{t:tab3}
\end{table}

In the case of DIS the corresponding expansion into a set of nine
covariants 
reads 
\begin{eqnarray}
\label{hmunuDIS}
\nonumber
H_{\mu \nu}&=& H_1 \left(g_{\mu \nu}-\frac{q_{\mu}
q_{\nu}}{q^2}\right) \\
\nonumber
& & + H_2 q^{-2} \hat{p}_{\mu} \hat{p}_{\nu}           \\
\nonumber
& & + H_3 q^{-2} \hat{p}_{\mu}^{\prime} \hat{p}_{\nu}^{\prime}      \\
\nonumber
& & + H_4 q^{-2} (\hat{p}_{\mu}\hat{p}_{\nu}^{\prime} +
\hat{p}_{\nu}\hat{p}_{\mu}^{\prime})                        \\
\nonumber
& & + H_5 q^{-2} (\hat{p}_{\mu} \hat{p}_{\nu}^{\prime} -
\hat{p}_{\nu} \hat{p}_{\mu}^{\prime})                     \\
& & + H_6 q^{-2} i \epsilon_{\mu\nu\alpha\beta} q^{\alpha} p^{\beta}    \\
\nonumber
& & + H_7 q^{-2} i \epsilon_{\mu\nu\alpha\beta} q^{\alpha} p^{\prime \beta} \\
\nonumber
& & + H_8 q^{-4} (\hat{p}_{\mu} {\tilde F}_{\nu} + \hat{p}_{\nu} {\tilde
F}_{\mu})                            \\
& & + H_9 q^{-4} (\hat{p}_{\mu}^{\prime} {\tilde F}_{\nu} + 
\hat{p}_{\nu}^{\prime} {\tilde F}_{\mu}),
\nonumber
\end{eqnarray}
where now $\tilde{F}_{\mu} = i \epsilon_{\mu\alpha\beta\gamma}
p^{\alpha}p^{\prime \beta}q^{\gamma}$. 
It is clear that the actual form of the parton level tensor $H_{\mu \nu}$
and the hard scattering structure functions $H_i$ depend on which of the
three quark-,
antiquark- and gluon-initiated cases are being discussed. If necessary, we
shall affix additional superfixes $q, \bar{q}$ and $g$, respectively, to
the parton level structure functions in order to differentiate between
the three cases.

For the Born term contribution and for the real parts of the one-loop
contributions the crossing can be done directly on the structure
function expansion (\ref{hmunu}) where one may have a reshuffling
of covariants in (\ref{hmunuDIS}) depending on whether one
is crossing to the 
quark-, antiquark- or the gluon-initiated case\footnote{The
relevant tables for the real structure functions $H_1,H_2,H_3,H_4,H_6$
and $H_7$ in $e^+ e^-$-annihilation can be found in \cite{KS}.}. 
In order to determine the absorptive
contributions to the 
$T$-odd structure functions $H_5,H_8$ and $H_9$ one has to go back to the
$e^+ e^-$ one-loop amplitude expressions in Table~\ref{t:tab2} and 
Table~\ref{t:tab3},
do the crossing, extract the
imaginary parts and finally fold them with the Born term amplitude.       
It is clear that $Im(N_7)$ does not contribute to the 
$T$-odd structure functions in this sequence of steps after folding with
the Born term amplitude since it is proportional to the Born term itself.
Our final expressions for all structure functions include spin and color
averaging factors for partons in the initial states. As indicated in 
Fig.~\ref{f:cross} the crossing of a fermion line brings in an overall
minus sign.

With all this we are now in a position to write down the $T$-odd hard
scattering 
structure functions $H_5^q, H_8^q$ and $H_9^q$ referring to the 
quark-initiated case of DIS. One uses 
the rules (\ref{qini}) and the results of Tables~\ref{t:tab1}, 
\ref{t:tab2}
and \ref{t:tab3}, together with the change of momenta defined in
(\ref{mDISq}) and the expansion (\ref{hmunuDIS}).
Note that the final antifermion line transforms to an initial fermion line
which implies an overall minus sign (see also Fig.~\ref{f:cross}). 

Averaging over the spin and color of the initial quark and summing over
the spins and colors of the final quarks and gluons 
one obtains 
the following results for the $T$-odd structure functions in the
quark-initiated case of DIS:
\begin{eqnarray}
\label{hq}
\nonumber
i\pi H_5^q&=& g^4 x^2 [ \frac{C_F N_C}{2} (1-\frac{2 x z}{(1-x)(1-z)}) 
                   -C_F (C_F-\frac{N_C}{2}) (1+\frac{2 z}{(1-x) (1-z)}  \\
\nonumber
& & {\rm \hspace{3.2in}}  +  \ln(z) \frac{2 z}{(1-x) (1-z)^2}) ],    \\ 
\nonumber
i\pi H_8^q&=&- g^4 \frac{x^3 (1-x-z)}{(1-x) (1-z)} [ \frac{C_F N_C}{2}
    +C_F (C_F-\frac{N_C}{2})(\frac{3-z}{1-z}+\ln(z)\frac{2}{(1-z)^2}) ], 
\\
i\pi H_9^q&=& g^4 \frac{x^3}{(1-x) (1-z)} [ \frac{C_F N_C}{2} 3
            +C_F (C_F-\frac{N_C}{2})
     (\frac{1-3 z}{1-z}+ \ln(z) \frac{2 (1-2z)}{(1-z)^2}) ],
\end{eqnarray}
where $x$ and $z$ are defined in (\ref{vardis}) and the color factors are
given by $C_F=4/3$ and $N_C=3$. 
The antiquark-initiated absorptive structure functions $H_i^{\bar{q}}$
($i=5,8,9$) can be obtained from the CP-invariance of the reaction. They
read
\begin{equation}
H_5^{\bar{q}}=H_5^q, {\rm\hspace{.4in}} H_8^{\bar{q}}=-H_8^q, 
{\rm\hspace{.4in}} H_9^{\bar{q}}=-H_9^q.
\end{equation}

In the gluon-initiated case one uses a corresponding crossing and colour
averaging procedure and obtains
\begin{eqnarray}
\label{hg}
\nonumber
i\pi H_5^g&=& g^4 \frac{x^2}{z (1-z)} [x-2 x z-\ln(z)\frac{1-x}{1-z}+
                 \ln(1-z) \frac{1-x}{z}] (C_F-\frac{N_C}{2}),        \\ 
\nonumber
i\pi H_8^g&=&- g^4 \frac{x^3}{z (1-z)} [2 (1-x)-
                  \frac{x+z-1}{z (1-z)}-\ln(z) \frac{x+z-1}{(1-z)^2}
                  -\ln(1-z) \frac{x-z-1}{z^2}] (C_F-\frac{N_C}{2}),  \\
i\pi H_9^g&=& g^4 \frac{x^3}{z (1-z)} [\frac{1-2 z}{z (1-z)}-
                         \ln(z)\frac{1}{(1-z)^2}+\ln(1-z)\frac{1}{z^2}] 
                                    (C_F-\frac{N_C}{2}).
\end{eqnarray}

As it can be noticed from (\ref{hg}) the QCD like contributions resulting
from the 3-gluon coupling vanish dynamically for the gluon-initiated case
(see also \cite{HHK1}).

We are now in the position to compare our results with those in
Ref.~\cite{HHK1}. In order to facilitate the comparison 
we contract the hadron tensor with the lepton
tensor $L_{\mu\nu}$. In the case of a pure vector coupling one has
\begin{equation}
\label{lmunu}
L^{\mu\nu}=2 [ k^{\mu} k'^{\nu} + k^{\nu} k'^{\mu} + \frac{q^2}{2}
   g_{\mu\nu} \mp i \epsilon^{\mu\nu\rho\sigma} k_{\rho} k'_{\sigma} ]
\end{equation}
where the upper(lower) sign holds for a purely left-(right-) handed
initial lepton. In the case of a left chiral ($V-A$) axial-vector charged
current without lepton polarization the form of (\ref{lmunu}) stays the
same except that one has only the upper ('minus') sign for the epsilon
term.

In this paper we are dealing with parton momenta only whereas in 
\cite{HHK1} the angular decay distributions are written down in terms of
hadron momenta.
We can nevertheless compare the angular structures of the two
approaches because in the parton model the parton momenta are assumed to
be collinear with the hadron momenta from which they originate or into
which they fragment. This allows us to meaningfully compare the angular
structure of the two approaches without having to set up the whole parton
model framework with its integrations over distribution and fragmentation
functions.


Calculating the hard scattering cross section we will have to take into
account an overall phase space factor
\begin{equation}
\label{ps}
        PS=\frac{1}{2^7 \pi^5} \frac{1}{ (-q^2)  }
\end{equation}
that multiplies the structure functions $H_i$, together with the
appropriate coupling constants.

\begin{figure}
  \centerline{\epsfig{file=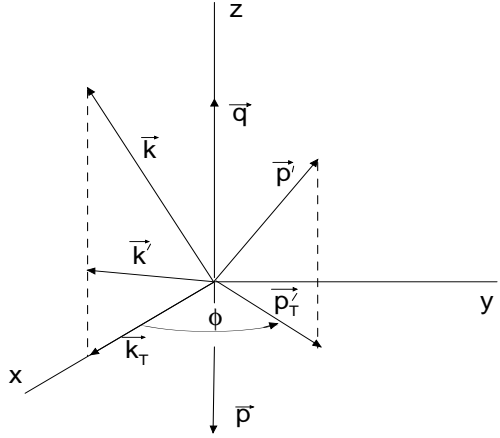,height=6cm,width=6.5cm,silent=}}
 \caption{ Three-momentum kinematics in the target rest frame in DIS. }
 \label{f:DIS}
\end{figure}

By contracting our parton level tensor $H_{\mu\nu}$ with the leptonic
tensor we obtain a result which has exactly the same form as the one in
\cite{HHK1} that stood for process variables. We find
\begin{equation}
\label{folding}
-q^{-2} L^{\mu\nu} H_{\mu\nu} = \frac{2}{y^2} \{A + B \cos\phi + 
                       C\cos2\phi + D \sin\phi + E \sin2\phi  \}
\end{equation}
with 
\begin{eqnarray}
\nonumber
A &=& 2 (1-y) F_L + [1 + (1-y)^2] F_T \pm y (2-y) F_3,      \\
\nonumber
B &=& \sqrt{1-y}\; [(2-y) F_4 \pm y F_5 ],                       \\
C &=& (1-y) F_6,                                          \\
\nonumber
D &=& \sqrt{1-y}\;[\pm y F_7 + (2-y) F_8 ],                    \\
\nonumber
E &=& (1-y) F_9,
\end{eqnarray}
where the $T$-odd coefficients $D$ and $E$ multiply the $T$-odd angular
correlation factors $\sin\phi$ and $\sin2\phi$, respectively. 
The upper(lower) sign is for the left-(right-) handed lepton beam.
The scaling variable $y$ is defined by
\[        y= \frac{qp}{kp}           \]
and is the same for parton and hadron kinematics. The same statement holds
true for the azimuthal angle $\phi$ between the lepton scattering plane 
and the hadron production plane (see Fig.\ref{f:DIS}).
As in \cite{HHK1}, 
we neglect the nonperturbative effects from the intrinsic transverse
momentum spread in the nucleon. 
Intrinsic transverse momentum effects have very
little influence on the $T$-odd asymmetries \cite{HHK1}.
We therefore assume that the azimuthal angle of the produced hadron is
identical to the azimuthal angle of the parton from which it originates.
From the definition of 
three-momenta drawn in Fig.\ref{f:DIS} one has
\[    \epsilon^{\alpha\beta\gamma\delta} 
      k_{\alpha} p_{\beta} {p'}_{\gamma} q_{\delta}=\frac{\kappa}{2x}
                             q^4\frac{\sqrt{1-y}}{y}\sin\phi,  \] 
where $\kappa=p'_T/(-q^2)^{1/2}$.

For the quark-initiated case one obtains the following expressions for
the angular coefficients $F_i,\, i=1,...,9$, in terms of the corresponding
nine structure functions:
\begin{eqnarray}
\label{fq}
\nonumber
F_L &=& - H_1^q + \frac{1}{4 x^2} H_2^q +
\frac{1}{4} \left ( z + \frac{\kappa^2}{z} \right )^2  H_3^q +
\frac{1}{2 x} \left ( z + \frac{\kappa^2}{z} \right ) H_4^q ,     \\
\nonumber
F_T &=& H_1^q + \frac{\kappa^2}{2} H_3^q ,                    \\
\nonumber
F_3 &=& \frac{1}{2} \left[ \left ( z + \frac{\kappa^2}{z} \right ) H_7^q
        + \frac{1}{x} H_6^q \right],                       \\
\nonumber
F_4 &=& - \kappa \left[ \left ( z + \frac{\kappa^2}{z} \right ) H_3^q
+\frac{1}{x} H_4^q \right] ,                                 \\
\nonumber
F_5 &=& - 2\kappa H_{7}^q                    \\
\nonumber
F_6 &=& \kappa^2 H_3^q ,              \\
F_7 &=& i \frac{\kappa}{x} H_{5}^q ,     \\
\nonumber
F_8 &=& - i \frac{\kappa}{2 x} \left[\frac{1}{x}H_{8}^q + \left ( z +
\frac{\kappa^2}{z} \right ) H_{9}^q \right] ,                  \\
\nonumber
F_9 &=& i  \frac{\kappa^2}{x} H_{9}^q.
\end{eqnarray}
The relations (\ref{fq}) agree with the corresponding
relations in \cite{HHK1} when 
$H_6^q$ and $H_7^q$ are
expressed by the $H_9^{HHK}, H_{10}^{HHK}$ as defined in \cite{HHK1} by
means
of Schouten's identity. In the quark-initiated case they read:
\begin{eqnarray}
\label{ttt}
\nonumber
H_6^q&=&-\frac{1}{4x} \left[ \left( z + \frac{\kappa^2}{z}\right )
         H_9^{HHK} 
        + x \left( z - \frac{\kappa^2}{z}\right )^2 H_{10}^{HHK} \right],
\\
H_7^q&=&\frac{1}{4x} \left[ \frac{1}{x} H_9^{HHK} + \left( z +
         \frac{\kappa^2}{z}  \right ) H_{10}^{HHK} \right] .
\end{eqnarray}

In the gluon-initiated case the relations between the angular coefficients
and the structure functions $H_i^g$ are identical to those in the
quark-initiated case.  \\
The $\kappa$-factor in (\ref{fq})-(\ref{ttt}) can be expressed as   
\[    \kappa=\sqrt{\frac{1-x}{x}z(1-z)}.  \]

Note that all the above results have been obtained for vector and
axial vector currents with unit strength, 
e.g. no numerical factors in the vertices are taken into account.
For the charged current case one has the factor
\[      \frac{g_w^4}{16} \frac{1}{M_w^4} \mid U_{ij} \mid ^2        \]
where $U_{ij}$ denotes the Kobayashi-Maskawa matrix element and
$g_w=e/\sin\theta_w$.
It is clear that the $P$-odd structure functions 
$F_8$ and $F_9$ vanish for purely electromagnetic interaction.

In \cite{HHK1} the dispersive structure functions
$F_L, F_T, F_3, F_4, F_6$
and $F_7$ have been calculated from the ${\cal O}(\alpha_s)$ Born term
contributions. The absorptive ${\cal O}(\alpha_s^2)$ structure functions  
$F_5, F_8$ and $F_9$ have been calculated in \cite{HHK1} directly in the
DIS channel by determining the appropriate cut contributions of the
one-loop diagrams in the DIS channel. The results on $F_5, F_8$ and $F_9$
in \cite{HHK1} agree with our results derived from crossing.

In a full ${\cal O}(\alpha_s^2)$ calculation of lepton-hadron correlations
in (2+1) jet production in DIS one would also need the one-loop
contributions to the set of dispersive structure functions \cite{BK}. As
remarked on earlier these can easily be obtained from the $e^+ e^-$
one-loop expressions listed in \cite{KS} and in this paper through
crossing.



For the sake of completeness we write down the result for
the hard scattering cross section in the case of the $Z$-boson exchange:
\begin{equation}
\label{zboson}
\frac{ k'_{0} p'_{0} d\hat{\sigma} }{d^3 k' d^3 p'} = \frac{{\cal P}}{2s}
\frac{2}{y^2} \{ {\cal A} + {\cal B} \cos\phi +
         {\cal C}\cos2\phi + {\cal D} \sin\phi + {\cal E} \sin2\phi  \}
\end{equation}
with
\begin{eqnarray}
\nonumber
{\cal A} &=& 2 (1-y) {\cal F}_L + [1 + (1-y)^2] {\cal F}_T \pm y (2-y)
{\cal F}_3,                                                \\
\nonumber
{\cal B} &=& \sqrt{1-y}\; [(2-y) {\cal F}_4 \pm y {\cal F}_5 ],       \\
{\cal C} &=& (1-y) {\cal F}_6,                                    \\
\nonumber
{\cal D} &=& \sqrt{1-y}\;[\pm y {\cal F}_7 + (2-y) {\cal F}_8 ],     \\
\nonumber
{\cal E} &=& (1-y) {\cal F}_9
\end{eqnarray} 
and ${\cal P}$ and $s$ are defined by
\[
 {\cal P} = \frac{1}{2^7 \pi^5} \frac{z}{q^4}
\delta(\kappa^2-\frac{1-x}{x}z(1-z)),
{\rm \hspace{.4in}}
s=(k+p)^2.
\]

For the parity conserving angular coefficients ${\cal F}^{PC}_i, \,\,
i=L,T,4,6,7$, which originate from the vector-vector and axial-axial
couplings, one has
\begin{equation}
{\cal F}^{PC}_i=F_{i}\left\{ e^4 e_l^2 e_q^2 +e^2 e_l e_q g_z^2
\frac{C_V^l\pm C_A^l}{2} C_V^q \mbox{Re}D +
g_z^4 (C_V^l\pm C_A^l)^2 \frac{(C_V^q)^2 +
(C_A^q)^2}{16} \mid D \mid^2 \right\}.
\end{equation}

For the parity violating angular coefficients ${\cal F}^{PV}_i, \,\,
i=3,5,8,9$, which originate from the vector-axial vector interference
contributions, we obtain
\begin{equation}
{\cal F}^{PV}_i=F_{i}\left\{ e^2 e_l e_q g_z^2 \frac{C_V^l\pm C_A^l}{2}
C_A^q \mbox{Re}D +
g_z^4 \frac{(C_V^l\pm C_A^l)^2}{8} C_V^q C_A^q \mid D \mid^2 \right\},
\end{equation}
where $e_l, e_q$ are lepton and quark charges, respectively. The upper
(lower) sign is for left- (right-) handed initial leptons, and 
\begin{eqnarray}
\nonumber
D&=&q^2 (q^2-m_Z^2+i m_Z \Gamma_Z)^{-1},                   \\
\nonumber
g_z&=&\frac{e}{\sin\theta_w \cos\theta_w},             \\
\nonumber
C_V^l&=&-\frac{1}{2}+2 \sin^2 \theta_w, {\rm \hspace{.2in}} 
C_A^l=-\frac{1}{2}       {\rm \hspace{.2in}} 
\mbox{for leptons with charge -1}, \\
\nonumber
C_V^q&=&\frac{1}{2} - \frac{4}{3} \sin^2 \theta_w, {\rm \hspace{.3in}} 
C_A^q=\frac{1}{2}  {\rm \hspace{.33in}} \mbox{for up quarks},       \\
C_V^q&=&-\frac{1}{2} + \frac{2}{3} \sin^2 \theta_w, {\rm \hspace{.2in}}
C_A^q=-\frac{1}{2} {\rm \hspace{.2in}} \mbox{for down quarks}.
\end{eqnarray}

The structure functions $F_i$ have to be taken from (\ref{fq})
for the quark- and gluon-initiated cases.
We stress that in (\ref{zboson}) terms proportional to Im$D$ have been  
dropped as they are of a higher order in the electroweak coupling.

\section{Crossing Results for Hadronic Structure Functions in DY}

In this section we present our results of crossing from the 
$e^+ e^-$-annihilation channel to the DY process. 
The changes in momenta resulting from 
crossing are illustrated on the right-hand side of 
Fig.~\ref{f:cross}. The calculation proceeds in 
analogy to that in DIS by following the rules defined in Sec.~II.

The decomposition of the $T$-odd parts of the hadron tensor 
$H_{\mu\nu}$ for the annihilation subprocess and both 
Compton scattering subprocesses is given by (in this section we
concentrate on the absorptive contributions)
\begin{eqnarray}
\label{hdy}
\nonumber
H_{\mu \nu}&=&
H_5 q^{-2} (\hat{p}_{a\mu} \hat{p}_{b\nu} - \hat{p}_{a\nu}\hat{p}_{b\mu}) \\
& & + H_8 q^{-4} (\hat{p}_{a\mu} F_{\nu} + \hat{p}_{a\nu} F_{\mu})     \\
\nonumber
& & + H_9 q^{-4} (\hat{p}_{b\mu} F_{\nu} + \hat{p}_{b\nu} F_{\mu}),
\end{eqnarray}
with
\begin{eqnarray}
\label{compton}
\nonumber
\hat{p}_{a\mu} &=& p_{a\mu} - \frac{p_a q}{q^2} q_{\mu},         \\
\nonumber
\hat{p}_{b\mu} &=& p_{b\mu} - \frac{p_b q}{q^2} q_{\mu},           \\
F_{\mu}&=& i \epsilon_{\mu\alpha\beta\gamma}
                p_a^{\alpha}p_b^{\beta}q^{\gamma}.
\end{eqnarray}

The results for our spin and color averaged T-odd structure
functions $H_5, H_8, H_9$ are given below. 
For the annihilation subprocess we obtain:
\begin{eqnarray}
\label{h589an}
-i\pi H_5^a &=&  g^4 C_F  \frac{(1+c)(x_a^2 -
x_b^2)}{4(1-x_a)(1-x_b)} 
\nonumber \\
&& +  g^4 \frac{C_F}{N_C} (C_F-\frac{N_C}{2})
    ( \frac{c x_a}{(1 - x_a)^2 (1 - x_b)} \ln\frac{c}{x_a} 
     + \frac{x_a ((x_a+x_b) (1-x_a x_b)+x_a^2+x_b^2)}{2 (1-x_a) (1-x_b)} )
\nonumber
\\
&&       - \{x_a\leftrightarrow x_b\},   \nonumber    \\
-i\pi H_8^a &=& -  g^4 C_F \frac{x_a (x_a^2-3
x_b^2)}{4(1-x_a)(1-x_b)} 
\nonumber \\ 
&& +  g^4 \frac{C_F}{N_C} (C_F-\frac{N_C}{2})
       x_a ( \frac{x_a (x_a-2 c)}{(1-x_a)^3 (1-x_b)} \ln\frac{c}{x_a}   
   - \frac{x_b^2}{(1-x_a) (1-x_b)^3} \ln\frac{c}{x_b}
\nonumber \\
&& + \frac{c(x_a^2-3x_b^2+2x_a+2x_b)-3x_a^2+x_b^2}{2(1-x_a)^2(1-x_b)^2} ), 
\nonumber     \\
-i\pi H_9^a &=& -i\pi H_8^a (x_a\leftrightarrow x_b),
\end{eqnarray}
where $c=x_a+x_b-x_a x_b$.

For the quark-initiated Compton scattering subprocess we find:
\begin{eqnarray}
\label{h589c}
-i\pi H_5^{C_q} &=& -\frac{g^4}{2} x_a^2 \frac{2 x_a x_b-c
(1+x_a)}{4c(1-x_a)}
\nonumber \\
&& + \frac{g^4}{2 N_C} (C_F-\frac{N_C}{2})
    x_a( x_a^2 (1-x_b)[ \frac{x_b}{c^2(1-x_a)}\ln(1-c)   
    + \frac{\ln(c/x_a)}{c(1-x_a)^2} ] \nonumber \\
&&+ \frac{c x_a (1+x_a)-2 x_a x_b (1+x_b)-4 x_b^3 (1-x_a)}{2 c (1-x_a)} ), 
\nonumber \\
-i\pi H_8^{C_q} &=& -\frac{g^4}{2}  x_a^3 \frac{x_a-2 x_b
(1+x_a)}{4c(1-x_a)} 
\nonumber \\ 
&& + \frac{g^4}{2 N_C} (C_F-\frac{N_C}{2})
       x_a^3( x_a[ \frac{x_b^2}{c^3 (1-x_a)} \ln(1-c)           
     + \frac{\ln(x_a/c)}{c (1-x_a)^3} ] \nonumber \\
&& + \frac{c x_a (x_a+4 x_b-2 x_a x_b+1)-
                       2 (x_a^2+x_a x_b^2-x_b^2)}{2 c^2 (1-x_a)^2} ),  
\nonumber \\
-i\pi H_9^{C_q} &=& \frac{g^4}{2}  \frac{3x_a^3 x_b}{4c (1-x_a)}
+
                                 \frac{g^4}{2 N_C}  (C_F-\frac{N_C}{2})
       x_a^2 x_b( x_a [ \frac{x_b (2 c-x_b)}{c^3 (1-x_a)}\ln(1-c) +
                         \frac{\ln(x_a/c)}{c (1-x_a)^3} ]   \nonumber \\
&&+             \frac{c x_a ((1-2 x_b)^2-3 x_a)+4 c x_b (1-x_b)+2 x_a
                   (x_a+x_b)}{2 c^2 (1-x_a)^2} ).
\end{eqnarray}
Note that we are using the same two colour structures as those in our $e^+
e^-$-annihilation expressions.

For the antiquark-initiated Compton subprocess one has
\begin{equation}
H_5^{C_{\bar{q}}}=H_5^{C_q}, {\rm\hspace{.4in}}
H_8^{C_{\bar{q}}}=-H_8^{C_q},
{\rm\hspace{.4in}} H_9^{C_{\bar{q}}}=-H_9^{C_q}.
\end{equation}

We consider only DY processes that proceed through
$W$-exchange. This is sufficient to compare our results with the results 
in Ref.~\cite{HHK3}. Following \cite{HHK3}, we consider the hard
scattering cross section
$d\hat{\sigma}/ d Q_T^2 d\!\cos\hat{\theta} d\!\cos\theta d\phi$.
For the $T$-odd part of the cross section we use the expansion
\begin{equation}
\label{expdy}
\frac{d\hat{\sigma}^{T-odd}}{d Q_T^2 d\!\cos\hat{\theta}
d\!\cos\theta d\phi}=
\sin\theta \sin\phi \,F_7 +
\sin2\theta \sin\phi \,F_8 +
\sin^2\theta \sin2\phi \,F_9 .
\end{equation}
\begin{figure}
 \centerline{\epsfig{file=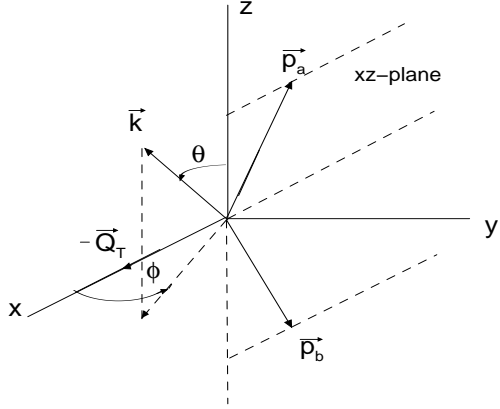,height=6cm,width=6.5cm,silent=}}
 \caption{ Three-momentum kinematics in the DY process for the
Collins-Soper frame. }
 \label{f:DY}
\end{figure}
The definition of angles is the same as in \cite{HHK3}.
The $W$+jet production is decribed by the transverse momentum $\vec{Q}_T$ 
of the jet and the
scattering angle $\hat{\theta}$ in the $W$+jet center of mass frame,
$\theta$ and $\phi$ are the polar and the azimuthal angle of the
lepton emerging from the decay $W \rightarrow l\nu$ in the Collins-Soper
frame \cite{CS} as shown in Fig.~\ref{f:DY}. A simple exercise in
particle and parton kinematics shows that the variables $\theta,
\phi$ and $\vec{Q}_T$ are identical in the hadron and parton processes.


The angular coefficients can be projected from the hadron tensor by means
of the Collins-Soper frame projectors
\begin{eqnarray}
\label{proj}
P_7^{\mu\nu}&=&\frac{i}{\sqrt{2}q^2} \frac{x_a x_b}{\sqrt{c(1-c)}}
            ( p_a^{\mu} p_b^{\nu} - p_a^{\nu} p_b^{\mu} ) ,
\nonumber \\
P_8^{\mu\nu}&=&-\frac{1}{\sqrt{2}q^4} \frac{x_a^2 x_b^2}{c\sqrt{1-c}}
                                                          [\frac{1}{x_b}
           (p_{a}^{\mu} \epsilon^{\nu\alpha\beta\gamma} p_{a\alpha}
            p_{b\beta} q_{\gamma} + \mu \leftrightarrow \nu) - \frac{1}{x_a}
           (p_{b}^{\mu} \epsilon^{\nu\alpha\beta\gamma} p_{a\alpha}
            p_{b\beta} q_{\gamma} + \mu \leftrightarrow \nu) ] ,
\nonumber   \\
P_9^{\mu\nu}&=&-\frac{\sqrt{c}}{q^4} \frac{x_a^2 x_b^2}{c(1-c)} [\frac{1}{x_b}
           (p_{a}^{\mu} \epsilon^{\nu\alpha\beta\gamma} p_{a\alpha}
            p_{b\beta} q_{\gamma} + \mu \leftrightarrow \nu) + \frac{1}{x_a}
           (p_{b}^{\mu} \epsilon^{\nu\alpha\beta\gamma} p_{a\alpha}   
            p_{b\beta} q_{\gamma} + \mu \leftrightarrow \nu) ] ,
\end{eqnarray}
which we use to extract the required components $F_i$ from 
the parton level tensor $H_{\mu\nu}$.

Then, taking into account the necessary numerical factors, we have the
following relations:
\begin{eqnarray}
\label{relation}
\nonumber
K^{-1} f_7  =  \frac{3}{\sqrt{2}} P_7^{\mu\nu} H_{\mu\nu}  &=&
                     i\pi\frac{3}{4}\frac{\sqrt{c(1-c)}}{x_a x_b}H_5 , \\
K^{-1} f_8  =  \frac{3}{2\sqrt{2}} P_8^{\mu\nu} H_{\mu\nu} &=&
      -i\pi\frac{3}{16}\frac{c\sqrt{1-c}}{x_a^2 x_b^2}(x_bH_8-x_aH_9) , \\
\nonumber
K^{-1} f_9  =  -\frac{3}{4} P_9^{\mu\nu} H_{\mu\nu} &=&
      i\pi\frac{3}{16}\frac{\sqrt{c}(1-c)}{x_a^2 x_b^2}(x_bH_8+x_aH_9),
\end{eqnarray}
where $K=8$ for the quark- and antiquark-initiated Compton
subprocesses and $K=3$ for the annihilation subprocess. The functions 
$f_7, f_8, f_9$ are related to the functions $F_7$, $F_8$ and $F_9$
appearing in eq.~(\ref{expdy}) and are defined 
in eqs.~(6) and (7) of \cite{HHK3}.

Substituting our expressions from (\ref{h589an}) and (\ref{h589c}) to
(\ref{relation})
we find complete agreement with the corresponding results of Eqs.~(8) 
and (9) of ref. \cite{HHK3}, except for the sign before the logarithmic
term of $f_{A9}$. This typo graphical error was corrected in the
footnote of \cite{HKY1} on p.~179.

\section{Summary and Outlook}

Starting from the known one-loop results for the quark-quark-gluon gauge
boson four-point function in the $e^+ e^-$-channel \cite{KS} we have used
analiticity and crossing to derive the absorptive parts of the same
four-point function in the DIS channel and in the DY process. Whereas the
imaginary parts of the one-loop four-point function generate a
nonmeasurable phase in $e^+ e^-$-annihilation one obtains measurable
phase effects in DIS and in the DY process leading to the nonvanishing of
$T$-odd observables which we have derived. We have compared our results
with the results of previous calculations where the absorptive parts in
DIS and in the DY process were calculated directly in the respective
channels.

In this paper we have mainly considered $T$-odd observables built from
triple products of the three-momenta of the respective processes. One can
also consider $T$-odd observables built from triple products involving
also spins. We have discussed such a triple product observable
involving the spin of the initial lepton in DIS. As shown in
Sec.~III the relevant $T$-odd observable is fed from the parity conserving
$T$-odd structure function $H_5$ given in this paper. 
The three absorptive invariant structure functions $H_5$,
$H_8$ and $H_9$ presented in this paper suffice 
to calculate any ${\cal O}(\alpha_s^2)$
contribution to $T$-odd observables as long as the spins of partons or
hadrons are summed over. If the $T$-odd triple product
involves the spins of hadrons participating in the process the relevant new
spin-dependent ${\cal O}(\alpha_s^2)$ contributions can be easily calculated
from the absorptive parts of the ${\cal O}(\alpha_s^{3/2})$ one-loop amplitudes
given in this paper. When one folds these with the respective Born term
expressions one has to keep the relevant spins unsummed. We hope to return to
the subject of spin-dependent $T$-odd contributions to DIS and the DY process in
the future.

\vskip 1.0cm
\acknowledgements
 
  We would like to thank K.~Hagiwara and K.~Hikasa for providing us with
  details of their calculations.
  We also thank T.~Brodkorb, L.~Br\"ucher and J.~Franzkowski for
  participating in the early stages of this work. 
  Z.M. thanks A.~Davydychev and H.S.~Do for discussions. 
  B.M. and Z.M. would like to thank the Institut f\"ur Physik, Universit{\"a}t 
  Mainz for hospitality and the BMBF, Germany, under contract 06MZ865,
  for support.
  This work was partially supported by the Ministry of Science and Technology
  of the Republic of Croatia under Contract No. 00980102.

\appendix
\section{The Triangle Anomaly in DIS}

\begin{figure}
 \centerline{\epsfig{file=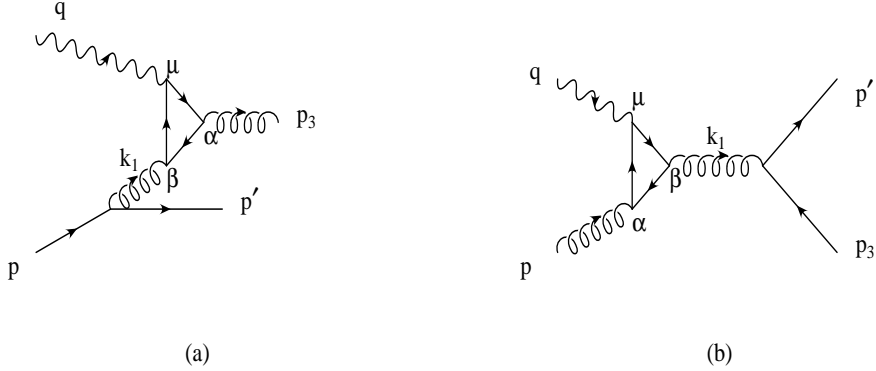,height=5.0cm,width=12.0cm,silent=}}
 \caption{ Graphs contributing to the $Zgg$ vertex: (a) quark-initiated
and (b) gluon-initiated case. Diagrams with counter-clock-wise quarks in
the loops are not shown.}
 \label{f:ANOM}
\end{figure}

In this Appendix we present the results for the axial-vector part of the
quark loop-induced 
$Zgg$ coupling contribution to DIS. 
To the best of our knowledge these
results have not been presented before. 
The relevant diagrams and assignments of momenta are shown in
Fig. \ref{f:ANOM}.
We use the results for the $Zgg$ vertex of
\cite{HKY1} and apply the above crossing procedure to
obtain new results. Following \cite{HKY1}, with momentum assignments
$Z(q)\rightarrow g(k_1)+g(k_2)$
we have the general
decomposition for the transition amplitude:
\begin{eqnarray}
\label{anom}
\nonumber
T_{\mu\alpha\beta}=f_1 (k_2^2 \epsilon_{\mu\alpha\beta\rho} k_1^{\rho}
+ k_{2\alpha} \epsilon_{\mu\beta\rho\sigma} k_{2\rho} k_{1\sigma}) +
f_2 (k_1^2 \epsilon_{\mu\alpha\beta\rho} k_2^{\rho} + 
k_{1\beta} \epsilon_{\mu\alpha\rho\sigma} k_{2\rho} k_{1\sigma}) +   \\
f_3 (k_1 + k_2)_{\mu} \epsilon_{\alpha\beta\rho\sigma} k_{2\rho}
k_{1\sigma}
+ f_4 (k_1 - k_2)_{\mu} \epsilon_{\alpha\beta\rho\sigma} k_{2\rho}
k_{1\sigma}),
\end{eqnarray}
where the functions $f_i$ depend on the Lorentz scalars $k_1^2, k_2^2$ and
$q^2=(k_1+k_2)^2$. As the gluon with momentum $k_2$ is on shell, $f_1$
does not
contribute. Due to the property $q_{\mu} L^{\mu\nu}=0$ in the massless
lepton limit $f_3$ does not contribute. The amplitude $f_4$ vanishes 
identically at the one loop level \cite{Hikasa} due to charge conjugation 
invariance. Technically speaking, the two contributions from the clockwise
and counter-clockwise quark flows in the quark loop cancel in $f_4$.


Thus, only the term proportional to $f_2$ remains. It has to be folded
with
the Born term amplitude. As it is well known, 
gauge invariance for the triangle graph with respect to the $Z$-boson
momenta is broken. This implies that the squared
matrix amplitude can be represented in the form of the decomposition
(\ref{hmunu}) plus additional terms proportional to $q_{\mu}$ or
$q_{\nu}$ which, however, do not contribute after folding with the
leptonic tensor.
Performing the necessary crossing and averaging over initial spins and colors, 
as defined in Sec.~3, one obtains the anomaly contribution to the DIS 
structure functions $H_i^q$.
For the quark-initiated case we have:

\begin{eqnarray}
\label{anomq}
H_1^q &=& \frac{4}{3} \frac{x+z}{z} q^2 {\rm Re}(f_2^q), \nonumber \\
H_2^q &=& \frac{8}{3} \frac{x(1+ x-z)}{z(1-z)} q^2 {\rm Re}(f_2^q),
\nonumber \\
H_3^q &=& \frac{8}{3} \frac{x(1-2 x)}{z(1-x)} q^2 {\rm Re}(f_2^q),
\nonumber \\
H_4^q &=& -\frac{4}{3} \frac{ (1-x)(2-z) - z (1-z)}{\kappa^2} q^2 
          {\rm Re}(f_2^q),                                 \nonumber \\
H_6^q &=& \frac{2}{3} \frac{(1-z)(1-x^2 + 2xz + (1-z)^2) + x(z-x)}{x
\kappa^2} q^2 {\rm Re}(f_2^q), \nonumber \\
H_7^q &=& -\frac{2}{3} \frac{ (1-z)( 2(1-x) (1-2x)+xz) + z^2(1-x)}{x
\kappa^2} q^2 {\rm Re}(f_2^q).
\end{eqnarray}

For the antiquark-initiated case results are the same except for
$H_6^{\bar{q}}=-H_6^q$ and
$H_7^{\bar{q}}=-H_7^q$. 
Note that there are no absorptive parts in the (anti)quark-initiated case
because $f_2^q$ does not have an imaginary part as we shell see later on.

For the gluon-initiated case we similarly obtain:
\begin{eqnarray}
\label{anomg}
H_1^g &=& -\frac{1-2 x}{2 (1-x)} q^2 {\rm Re}(f_2^g), \nonumber \\
H_2^g &=& \frac{z(1+ x-z)}{\kappa^2} q^2 {\rm Re}(f_2^g), \nonumber \\
H_3^g &=& \frac{1}{\kappa^2} q^2 {\rm Re}(f_2^g), \nonumber \\
H_4^g &=& -\frac{ 1-x+2 x z}{2\kappa^2} q^2 {\rm Re}(f_2^g), \nonumber \\
H_5^g &=& i \frac{ x(1-2 z)}{2 z (1-z)} q^2 {\rm Im}(f_2^g), \nonumber \\
H_6^g &=& -\frac{(1-z)( 2x - (1-z)^2 - (z-x)^2 ) - xz(1-x)}{4x\kappa^2} 
                                 q^2 {\rm Re}(f_2^g),         \nonumber \\
H_7^g &=& -\frac{ 1-x - 2 z (1-z) (1+ 2 x) }{4 x \kappa^2} q^2 {\rm
                               Re}(f_2^g),                   \nonumber \\
H_8^g &=& i \frac{x z}{\kappa^2} q^2 {\rm Im}(f_2^g),     \nonumber \\
H_9^g &=& i \frac{x (1-2 z)}{\kappa^2} q^2 {\rm Im}(f_2^g).
\end{eqnarray}

The functions $f_2^q$ and $f_2^g$ can be easily written down using
eqs.~(2.5)-(2.9) of \cite{HKY1} for the DIS region where $q^2<0$. For the
(anti)quark-initiated case one also has $k_1^2<0$.
Then we arrive at the following expression for $f_2^q$:
\begin{eqnarray}
\label{f2q}
\nonumber
f(w,r) \rightarrow f_2^q
=-\frac{1}{2 \pi^2 q^2} \frac{x^2}{(x-z)^2} \left[ \frac{1}{r} \left\{
       \ln^{2}(\sqrt{1-rw} + \sqrt{-rw}) - \ln^{2}(\sqrt{1-r} + \sqrt{-r})
\right\} -
\right .                \\
 \left .
        2 \left\{ \sqrt{\frac{1-rw}{-rw}} \ln(\sqrt{1-rw} + \sqrt{-rw}) -
       \sqrt{\frac{1-r}{-r}} \ln(\sqrt{1-r} + \sqrt{-r}) \right\} + \ln w 
                       \right],
\end{eqnarray}
with $w=k_1^2/q^2$ and $r=q^2/4m^2$ and where all quark masses are set to
zero except for the top quark mass denoted by $m$.
This approximation is valid at high enough energies when only
the mass of the top quark is important. 
Then the contributions from the first two generations of "light" quarks 
cancel out between the up($u,c$)- and down($d,s$)-quarks and only $b$- and
$t$-quarks contribute to the above functions. 

One can see that $f_2^q$ is a real function, and therefore the $T$-odd
structure functions in the (anti)quark-initiated case of DIS do not
receive any anomaly contribution.

However, for the gluon initiated case we have $k_1^2>0$, and, clearly,
there will be nonvanishing imaginary contribution from the triangle
diagram. We have to separately consider the two regions below and above
top threshold. 
Below top threshold with $q^2<0, \,\, 0<k_1^2<4 m^2 \,\, (0<rw<1)$ we get:
\begin{eqnarray}
\label{f2g1}
{\rm Re} f_2^g &=& \frac{x^2}{2 \pi^2 q^2}  \left[
\frac{1}{r} \left\{ (\sin^{-1} \sqrt{rw})^2 + \ln^{2}(\sqrt{1-r} + \sqrt{-r})
\right\} +
\right .                \\   \nonumber
&&   \left .
2 \left\{ \sqrt{\frac{1-rw}{rw}} \sin^{-1} \sqrt{rw} -
\sqrt{\frac{1-r}{-r}} \ln(\sqrt{1-r} + \sqrt{-r}) \right\} - \ln \mid w \mid
                       \right],                   \\
\label{f2g11}
{\rm Im} f_2^g &=& \pi
\end{eqnarray}
Since one is below top quark threshold there is only an imaginary part
coming from the $b$-quark. The contributions of the ($u,d$) and ($c,s$)
quarks cancel pairwise in the real and in the imaginary parts. The
$b$-quark contribution in the real part is proportional to the last term
in (\ref{f2g1}).

Above top threshold with 
$q^2<0,\,\, k_1^2\geq 4 m^2\,\, (rw\geq 1)$ we have
a $k_1^2$-dependent imaginary part:
\begin{eqnarray}
\label{f2g2}
{\rm Re} f_2^g &=& \frac{x^2}{2 \pi^2 q^2}  \left[
\frac{1}{r} \left\{ \frac{\pi^2}{4} - \ln^{2}(\sqrt{rw} + \sqrt{rw-1}) +
\ln^{2}(\sqrt{1-r} + \sqrt{-r})
\right\} +
\right .                \\   \nonumber
&&   \left .
2 \left\{ \sqrt{\frac{rw-1}{rw}} \ln(\sqrt{rw} + \sqrt{rw-1}) -
\sqrt{\frac{1-r}{-r}} \ln(\sqrt{1-r} + \sqrt{-r}) \right\} - \ln \mid w \mid
                       \right],                   \\
\label{f2g22}
{\rm Im} f_2^g &=& \pi \left[ 1 + \frac{1}{r} \ln(\sqrt{rw} + \sqrt{rw-1}) -
                    \sqrt{\frac{rw-1}{rw}} \right].
\end{eqnarray}

The trivial imaginary term in (\ref{f2g11}) comes entirely from the
"light" $b$-quark contribution. It is also present in (\ref{f2g22}),
which now also has an imaginary part due to the $t$-quark because now
one is above top threshold. 
In the limit of zero $t$-quark mass the contribution
from the triangle graph should vanish. In this case (\ref{f2g1}) and
(\ref{f2g11}) would
be obviously absent and, as one can easily see, the real and imaginary
contributions in (\ref{f2g2}) and (\ref{f2g22})
vanish in that limit, serving as a partial check of the correctness of the
above expressions.

\end{document}